# Slowing down of the relaxational dynamics at the ferroelectric phase transition in one-dimensional $(TMTTF)_2AsF_6$

D. Starešinić[a,b], K. Biljaković[a], P. Lunkenheimer[b], A. Loidl[b]

[a]*Institute of Physics, Zagreb, Croatia*
[b]*Experimental Physics V, Center for Electronic Correlations and Magnetism, University of Augsburg, Germany*

**Abstract**

We present measurements of the dielectric response of quasi one-dimensional $(TMTTF)_2AsF_6$ in a wide temperature and frequency range. We provide a thorough characterization of the relaxational dynamics observed close to the ferroelectric-like transition at $T_c = 100$ K. Our measurements, extending up to 100 MHz, reveal a continuous slowing down of the mean relaxation time when approaching $T_c$ from high as well as from low temperatures. The simultaneous critical rise of dielectric constant and relaxation time point to an explanation of the transition in terms of a classic ferroelectric scenario.

## 1. Introduction

Low dimensional electronic systems, of which the organic charge transfer salts $(TM)_2X$ [1,2] are paramount examples, serve as model systems for the investigation of strong correlations. They are formed of stacks of the organic molecules tetramethyltetrathiofulvalene (TM=TMTTF) [1] or tetramethyltetraselenafulvalene (TM=TMTSF) [2] with X=$PF_6$, $AsF_6$, $SbF_6$, Br, SCN, $ClO_4$, $ReO_4$, etc. as counterions, which occupy loose cavities formed by the methyl groups of the organic molecules. A variety of strongly correlated electronic ground states is observed at low temperatures including superconductivity, spin and charge density waves (SDW/CDW), Neel antiferromagnetism (AFM) and Mott-Hubbard (or Wigner-type) charge localization, as well as a spin-Peierls ground state. They can be presented by a unified phase diagram depending on temperature and external or chemical pressure [3]. Systems with non-centrosymmetric (NCS) anions (X=SCN, $ClO_4$, $ReO_4$, etc.) also exhibit anion ordering at finite temperature.

Only in systems with NCS anions, their ordering at finite temperature seems to affect the properties of $(TM)_2X$ systems [4]. However, recently it has been shown [5,6,7,8] that even the centrosymmetric (CS) anions in the TMTTF family undergo a phase transition that involves a uniform displacement of the anion sublattice relative to the cation chains, which in turn induces electronic charge disproportionation due to the lifting of inversion symmetry. Thus at low temperatures these materials exhibit a combination of conventional displacive ionic and electronic ferroelectricity. This transition bears the characteristics of ferroelectric (FE) order in dielectric spectroscopy, as deduced from the Curie-Weiss like divergence of the dielectric constant [5,6,7,8]. As optical measurements [9] gave no evidence for a FE soft-mode expected for a second-order displacive transition, Brazovskii et al. proposed [10] that a soft mode in TMTTF probably is accessible at low frequencies, consistent with a non negligible low-frequency dispersion [6,7]. A theoretical proposal [5,10] relates the FE transition to the change of the average phase in the charge channel of the Luttinger-liquid ground state, which is balanced by the spatial shift of the anions, leading to a displacive transition of second order. However, it also allows for the existence of domains of different polarization, which could result in dissipative low frequency dynamics.

In [7] the low frequency dielectric response of $(TMTTF)_2PF_6$ is dissipative around the FE transition at $T_c$=60 K and the mean relaxation time follows an activated temperature dependence. The activation energy is smaller below than above the FE transition. It has been argued that this behavior is related to the FE domain dynamics. The data for $(TMTTF)_2AsF_6$ [6] on the other hand are less dispersive around $T_c$=100 K and indicate that the relevant dynamics is located at higher frequencies. It seemed interesting to extend the frequency range to higher frequencies, to make a detailed analysis in order to capture the dynamics in $(TMTTF)_2AsF_6$ and compare it to that in $(TMTTF)_2PF_6$.

## 2. Experiment

Altogether ten $(TMTTF)_2AsF_6$ samples of typical lengths of 2-4 mm and cross-sections of $10^{-4}$ – $10^{-3}$ $mm^2$ have been measured in a wide frequency (1 Hz – 3 GHz) and temperature range (20 K - 300 K). Three different devices have been used, the HP4284 impedance analyzer (frequency range 25 Hz – 1 MHz) in pseudo four-point configuration and the HP4991 (1 MHz – 3 GHz) and Agilent 4294 (40 Hz – 110 MHz) impedance analyzers in coaxial reflection configuration [11]. The ac amplitude was 5 – 30 mV, well within the linear dielectric response regime, particularly around $T_c$. All of our samples suffered from cracks on the initial cooling from room temperature down to about 200 K, which reduced the room temperature conductivity after the measurements by almost one order of magnitude. However, after this initial "annealing" all samples exhibited a very similar temperature dependence, particularly with a well defined change of slope at about 100 K.

## 3. Results

The data presented in this paper are obtained on a sample for which the features near the FE transition are most distinguished, however, they are qualitatively the same as for the other samples. The temperature dependence of the real part of the complex conductivity σ' is presented in Figure 1 for selected frequencies. These data have been taken on heating, where no additional cracks appeared. A significant frequency dependence is observed already below 200 K. For the lowest frequencies, where σ'(T) provides a good estimate for the dc conductivity, a change of slope at $T_c$ is observed [6]. For the higher frequencies, peaks or shoulders show up in the vicinity of $T_c$, which seem to shift towards lower temperatures with decreasing frequency. As the dielectric loss ε'' is proportional to σ'/ν, this finding is typical for relaxational processes as observed, e.g., in relaxor ferroelectrics [ 12 ], the peaks in ε'' (respectively σ') arising when the condition $\omega\tau(T) = 1$ is fulfilled, with τ being the mean relaxation time and $\omega = 2\pi\nu$.

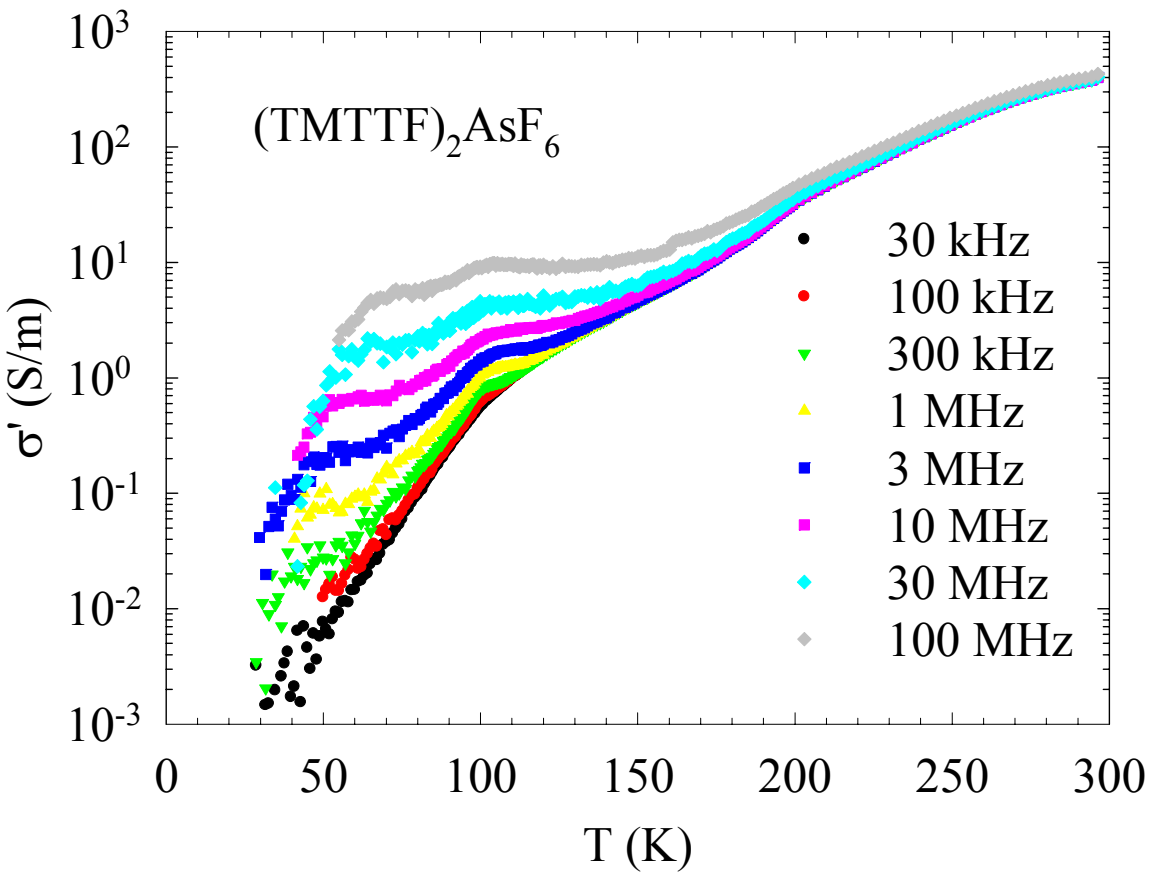

Figure 1. Temperature dependence of the real part of the complex conductivity σ' at selected frequencies.

The temperature dependence of the real part of the dielectric function ε' is presented in Figure 2 for selected frequencies. A sharp peak at 100 K is observed at low frequencies while for higher frequencies only a shoulder is present. The peak is superposed by a substantial background increasing with temperature, which may be due to interfacial effects at cracks that have formed in the sample during cooling. As shown in the inset of Figure 2, ε'(T) for the lowest frequency, which we identify with the static susceptibility Δε, follows a Curie-Weiss law above $T_c$, with substantial rounding at $T_c$. Below the transition, the slope changes sign but is similar in absolute values, not in accordance with predictions of a Curie-Weiss behavior in conventional second-order FE phase transitions where a slope ratio of 1:2 is expected (cf. dashed line in the inset of Fig. 2). This behavior could result from a disorder-induced smearing out of the phase transition

The dispersion of ε', revealed in Figure 2, clearly indicates relaxational behavior with the points of inflection below $T_c$ shifting towards higher temperatures with increasing measuring frequency. Relaxation processes lead to a steplike decrease of ε'(T) under decreasing temperature. In $(TMTTF)_2AsF_6$ these relaxation steps are superimposed on a strong Curie-Weiss temperature dependence of the static susceptibility, which leads to peaks in ε' (T) shifting with frequency. Remarkably, such a behavior is considered a hallmark feature of the so-called relaxor ferroelectrics, where the typical strong dispersion effects often are ascribed to the freezing-in of ferroelectric clusters [12]. However, already a closer inspection of Figure 2 reveals, that for the highest frequencies, the peaks in ε'(T) seem to cease shifting or even reverse the direction of the shift.

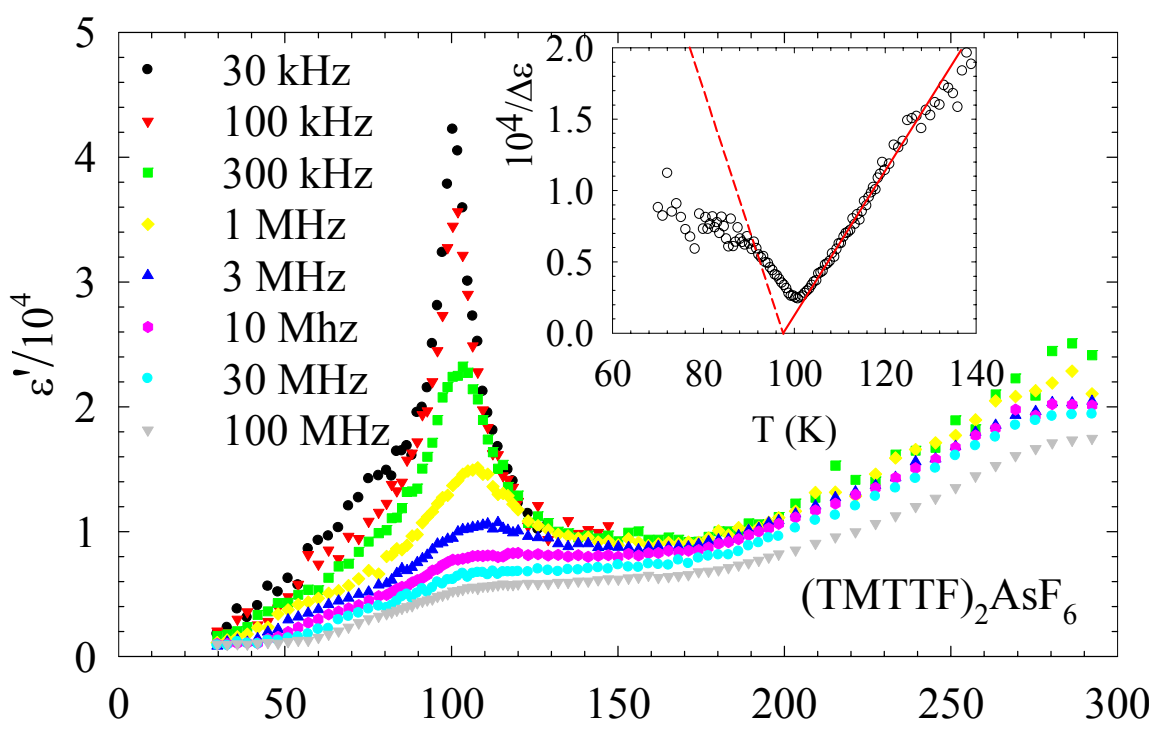

Figure 2. Temperature dependence of the real part of the complex dielectric constant ε' at selected frequencies. The inset shows the inverse static susceptibility. The solid line indicates Curie-Weiss behavior; below $T_c$ the expected behavior for a conventional second-order ferroelectric phase transition (slope ratio 2) is indicated by the dashed line.

The most significant information on relaxational dynamics can be gained from frequency-dependent plots of the permittivity. Thus the frequency dependent dielectric loss ε''(ν) has been calculated from the conductivity σ'(ν) after subtraction of the dc conductivity $\sigma_{dc}$ using $\varepsilon''(\nu) = (\sigma'(\nu) - \sigma_{dc}) / (\omega\varepsilon_0)$ with $\varepsilon_0$ the permittivity of free space. In Figure 3, ε''(ν) is presented together with ε'(ν) for several temperatures around $T_c$. The slowing down of the relaxation dynamics with temperature decreasing towards 100 K (Figure 3a) becomes obvious from the shift of the relaxation step in ε'(ν) and the peak in ε''(ν) as is typical for relaxational processes in most materials, including relaxor ferroelectrics. However, as revealed in Figure 3b, under further cooling below $T_c$, the relaxation speeds up again! To obtain more quantitative information on this finding, the spectra at various temperatures were fitted to the phenomenological Cole-Cole function:

$$\varepsilon' - i\varepsilon'' = \frac{\Delta\varepsilon}{1-(i\omega\tau)^{1-\alpha}} + \varepsilon_\infty \qquad (1)$$

Here Δε is the relaxation strength, $\varepsilon_\infty$ the high-frequency limit of the dielectric constant, and α a width parameter. Compared to single-exponential Debye behavior corresponding to $\alpha = 0$, values of $0 < \alpha < 1$ result in broadened loss peaks. Such deviations from Debye behavior are commonly ascribed to a distribution of relaxation times arising from disorder. For the spectra shown in Fig. 3, the lines show the results of fits with eq. (1), performed simultaneously for ε' and ε'', leading to a reasonable description of the experimental data.

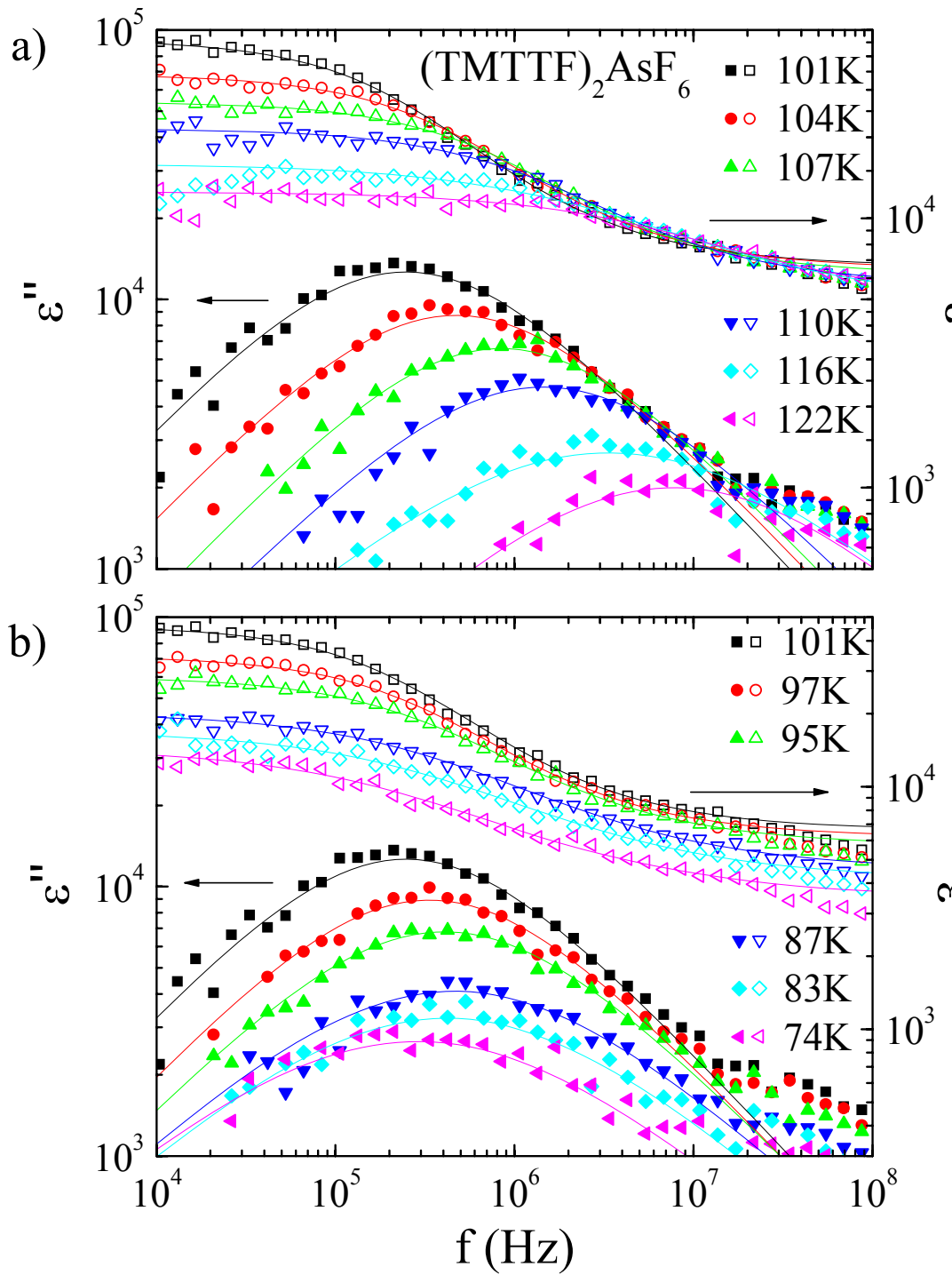

Figure 3. Frequency dependence of the dielectric constant ε' (open circles) and loss ε'' (closed circles) at several temperatures above (a) and below (b) $T_c$. The solid lines represent fits to the Cole-Cole expression, eq. (1).

The temperature dependence of the relaxation time τ resulting from the fits, which is related to the position of the maximum in ε''(ν) via $\tau = 1/(2\pi\nu_{max})$, is shown in Figure 4 in an Arrhenius representation. It increases as the temperature decreases towards $T_c$, but below $T_c$ initially it *decreases* significantly before starting to increase again. Thus, the relaxation slows down near $T_c$ and becomes faster again immediately below. The increase of τ towards $T_c$ can be well described by the Vogel-Fulcher-Tammann law (dashed line in Figure 4) typical for relaxor ferroelectrics [12,13]:

$$\tau(T) = \tau_0 e^{\frac{E_a}{T-T_0}} \qquad (2),$$

with $\tau_0 = 2\cdot10^{-9}$ s, $E_a = 189$ K and $T_0 = 77$ K. However, in relaxor ferroelectrics usually there is no anomaly of τ(T) at $T_c$. In principle, the rounding of the transition, the strong dispersion effects, and the polydispersive behavior of the relaxation process are hallmark features of glasslike freezing as in relaxor ferroelectrics. In the present case of $(TMTTF)_2AsF_6$, one could speculate that long-range FE order is suppressed by the one-dimensionality of the system.

However, the temperature dependence of the dielectric constant and the critical enhancement of the mean relaxation rate at $T_c$ may also be explained in terms of a classical FE behavior derived from a soft-mode behavior and including order-disorder phenomena. In this case the static dielectric constant reveals the classical behavior of a second-order FE phase transition (however with some quantitative deviations; see inset of Fig. 2) and the relaxation time enhancement close to $T_c$ can be interpreted as critical behavior. It has to be noted that in this case the soft mode behavior is dominated by the critical temperature dependence of the damping [14].

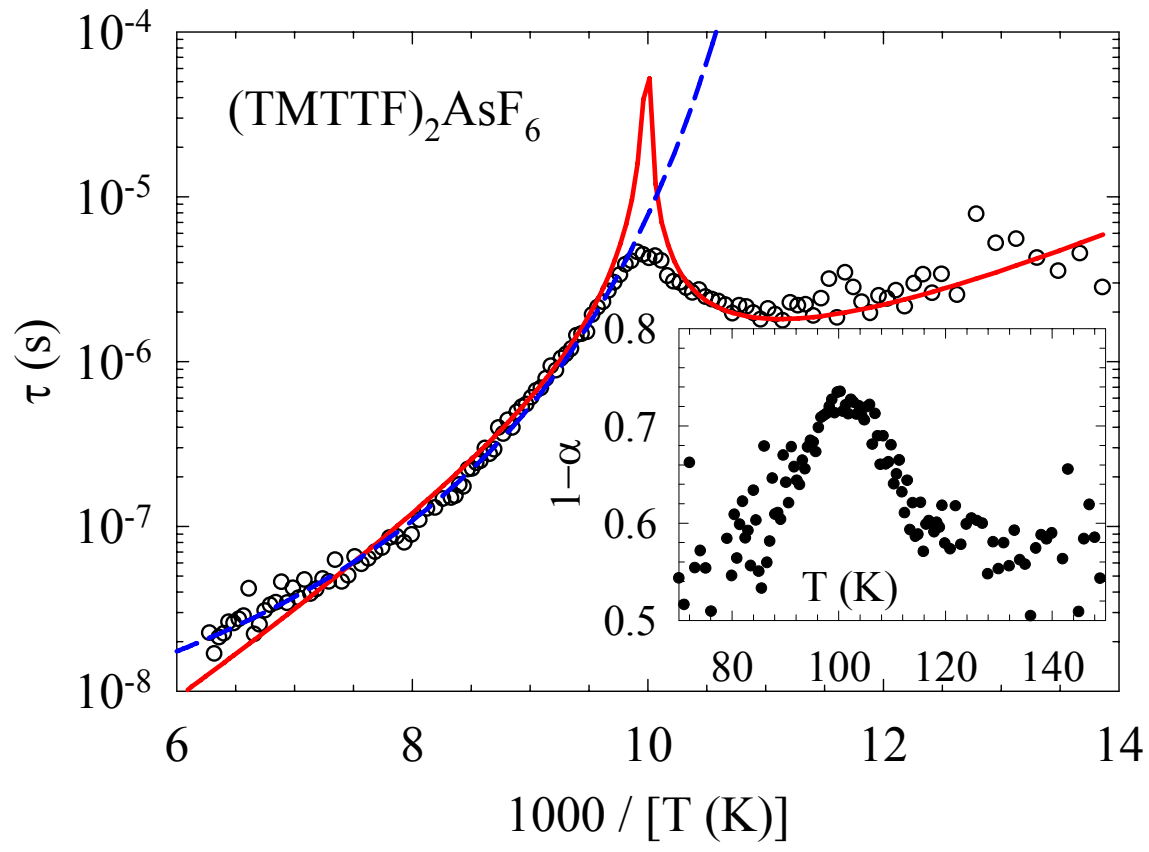

Figure 4. Temperature dependence of the mean relaxation time τ. The dashed line represents a fit to the expression (2), and the solid line a fit to the expression (3). The inset shows the temperature dependence of the width parameter of the Cole-Cole function, 1-α.

The solid line in Figure 4 is a fit to the expression

$$\tau(T) = \frac{\tau_0 T e^{\Delta/T}}{A|T-T_c|}; \quad A = \begin{cases} 1 & T > T_c \\ 2 & T < T_c \end{cases} \qquad (3),$$

which, founded on the theoretical treatment of a order-disorder FE transition in the case of a shallow double well potential [ 15 ], phenomenologically combines a thermally activated temperature dependence and critical behavior close to $T_c$. The activation energy $\Delta$ obtained from the fit is about 950 K, of the same order as the activation energy of $\Delta_\sigma$=700 K obtained from the dc conductivity below $T_c$. However, with $1 \cdot 10^{-11}$ s, the value of $\tau_0$ seems rather low.

In the inset, the temperature dependence of the width parameter 1-$\alpha$ of the Cole-Cole equation, which parameterizes the distribution width, is shown. In the vicinity of the FE transition, the width becomes narrower, changing its character from polydispersive to rather monodispersive. Thus, in the critical regime close to $T_c$, the polar fluctuations are governed by one unique relaxation time.

## 4. Discussion

Concerning the character of the FE transition, our results resemble more closely those in $(TMTTF)_2PF_6$ [7] than the ones reported originally for $(TMTTF)_2AsF_6$ [6] where, however, no information on the temperature dependence of the relaxation time was given. Without the local maximum at $T_c$, our $\tau$(T) may be described just as in [7], namely by thermally activated behavior with a decrease of the effective energy barrier below $T_c$. Vice versa, it can be argued that in $(TMTTF)_2PF_6$, while the tendency exists, the maximum of $\tau$(T) at $T_c$ is not observed because the transition is much more smeared out compared to $(TMTTF)_2AsF_6$.

Soft ferroelectric modes, i.e. dissipative processes that "soften" (become slower) when approaching the FE transition are well known for order-disorder FE systems and assumed to be directly connected to the displacive ferroelectric phase transition. The $\tau$(T) dependence in $(TMTTF)_2AsF_6$ could be understood as a consequence of the collective nature of dipolar excitations. The FE transition has been described [5,10] as the spatial ordering of the phase in the charge channel of the Luttinger liquid formed in decoupled TMTTF stacks. Two competing dimerization potentials, built-in bond dimerization and spontaneous site dimerization, lead to the existence of two distinct stable values of the charge density phase below $T_c$. The energy barrier between two possible states is equal to the gap in the electronic spectrum, which explains why the activation energy obtained from $\tau$(T) is so close to the one obtained from $\sigma_{dc}$(T). The microscopic relaxation time $\tau_0$, on the other hand, is determined by the dynamics of topological excitations (solitons) which change the phase between two stable values. As the effective mass of solitons in quasi one-dimensional systems [16] can be up to 2 orders of magnitude higher than the electron mass, this may explain the high value of $\tau_0$.

However, we are reluctant to unequivocally ascribe the observed relaxation process in $(TMTTF)_2AsF_6$ to such a mode, as their typical relaxation rates are much higher than in the present case, usually featuring non-distributed (Debye) frequency dependence and only in some cases reaching values as low as GHz [17,18,19]. The strong dispersion effects already at radio frequencies may be due to the low-dimensional character of the system under investigation: It has been shown [20] that in quasi one-dimensional systems the critical dynamics exhibits a stronger slowing down due to the anisotropy of the exchange. An alternative explanation arises when considering that dissipative relaxation processes in the kHz - MHz range with an activated or Vogel-Fulcher-Tammann temperature dependence of $\tau$ and a distribution of relaxation times are well established features of relaxor ferroelectrics [12,13].

## 5. Summary

We have measured the dielectric response of the quasi one-dimensional organic conductor $(TMTTF)_2AsF_6$ in a wide frequency and temperature range. The temperature dependence of the dielectric constant (as previously reported in [5]) is consistent with the existence of a ferroelectric phase transition at $T_c$ = 100 K. Dissipative effects exhibiting the clear signature of relaxational processes show up in the vicinity of the phase transition. The temperature dependence of the relaxation time shows a slowing down when approaching the transition from above and below $T_c$. It can be described by an expression suitable for order-disorder transitions involving shallow double well potentials. However, an interpretation in terms of a relaxor ferroelectric state also seems possible. Close to $T_c$, the relaxation almost becomes monodispersive. The dispersion occurring at unusually low frequencies in the TMTTF family can be explained by the one-dimensional character of the system.

## Acknowledgements

This work was partly supported by the Deutsche Forschungsgemeinschaft via the Sonderforschungsbereich 484 and partly by the BMBF via VDI/EKM (13N6917-B). D. Starešinić has been supported by the Alexander von Humboldt Foundation.